\documentclass[twocolumn]{jpsj3}

\usepackage{color}
\usepackage{bm}

\title{%
Local Density of States in a Helical Tomonaga-Luttinger Liquid
of Loop and Josephson Junction Geometries
}

\author{%
Yositake Takane
}

\inst{%
Department of Quantum Matter, Graduate School of Advanced Sciences of Matter,\\
Hiroshima University, Higashihiroshima, Hiroshima 739-8530, Japan
}

\recdate{ \hspace{50mm} }

\abst{%
The local density of states (LDOS) in a one-dimensional helical channel
of finite length
is studied within a Tomonaga-Luttinger model at zero temperature.
Two particular cases
of loop and Josephson junction geometries are considered.
The LDOS, as a function of energy $\omega$ measured from the Fermi level,
consists of equally spaced spikes of the $\delta$-function type,
and electron-electron interactions modify their relative height.
It is shown that, in the loop geometry, the height of spikes decreases as
$\omega \to 0$ everywhere in the system.
It is also shown that, in the Josephson junction, the behavior of
the LDOS significantly depends on the spatial position.
At the end points of the junction, the height increases as $\omega \to 0$
and its variation is more pronounced than that in the loop case.
Away from the end points, the height of spikes shows a non-monotonic
$\omega$-dependence, which disappears in the long-junction limit.
}

\begin{document}
\sloppy
\maketitle

\section{Introduction}

The existence of a one-dimensional (1D) edge channel
with a linear energy dispersion is the most notable property of
quantum spin Hall insulators.~\cite{kane,onoda,bernevig,qi,murakami}
This 1D channel is called helical as it hosts up-spin and down-spin
electrons moving in opposite directions.
As a consequence of the time-reversal invariance of
quantum spin Hall insulators, these two branches are
connected by time-reversal symmetry.
This forbids impurity-induced single-particle backward scattering.
The absence of backward scattering is a remarkable property of
the 1D helical channel allowing it to be free from Anderson localization.

The electron-electron interaction effect on the 1D helical channel
has been studied in Refs.~\citen{wu} and \citen{xu}.
It is shown that the system remains gapless even in the presence of disorder
unless interactions are extraordinarily strong,
and is described by the concept of a Tomonaga-Luttinger
liquid.~\cite{tomonaga,luttinger,mattis,luther,suzumura1,haldane}
Since this helical liquid has only one time-reversal invariant pair of
up-spin and down-spin branches moving in opposite directions,
its description is simpler than that of
an ordinary spin-full Tomonaga-Luttinger liquid.
Hence, the 1D helical channel of quantum spin Hall insulators
can be regarded as an ideal platform
to examine the characteristic behavior of the Tomonaga-Luttinger liquid.

Although the 1D helical channel is most naturally realized on the edge of
quantum spin Hall insulators, it has been demonstrated that
the step defect on a certain surface of weak topological insulators
can also host an equivalent 1D channel.~\cite{yoshimura}
This channel inevitably forms a closed loop structure
as discussed in Ref.~\citen{yoshimura},
indicating that a Tomonaga-Luttinger liquid in the loop
geometry~\cite{loss,schmeltzer,fujimoto,giamarchi,odintsov,pletyukhov}
can be realized on a surface of weak topological insulators.
The system similar to this is also realized in the Josephson
junction~\cite{maslov,takane1,affleck,takane2,caux,saha,barbarino}
of a Tomonaga-Luttinger liquid, which can be made by depositing
two superconductors on a quantum spin Hall insulator.
In the Josephson junction, electrons near the Fermi level is
effectively confined in the finite region between the superconductors.
In such systems, how does the finiteness of an effective system length
affect the behavior of the 1D helical channel?

In this paper, we focus on the local density of states
in the 1D helical channel at zero temperature
and study the finite-size effect on it
combined with the effect of interactions
within the Tomonaga-Luttinger liquid theory.
We consider the two cases of the loop and Josephson junction geometries.
If the local density of states is plotted
as a function of the energy $\omega$ (measured from the Fermi level),
it consists of equally spaced spikes of the $\delta$-function type,
reflecting the finiteness of system length,~\cite{comment} and
the electron-electron interactions modify the height of these spikes.
It is shown that the $\omega$-dependence of peak height is determined by
the correlation exponent $K$ characterizing the strength of interactions.
In the loop geometry case, the height of spikes decreases as $\omega \to 0$
everywhere in the loop.
It is also shown that, in the Josephson junction case, the behavior of
the local density of states significantly depends on
the spatial position in the junction between two superconductors.
At the end points of the junction, the height of spikes increases
as $\omega \to 0$ in contrast to the loop case.
Furthermore, its variation as a function of $\omega$ is
much more pronounced than that in the loop case.
Away from the end points, the height of spikes shows a non-monotonic
$\omega$-dependence,
which disappears in the limit where the system length is large enough.

In the next section, we describe a weakly interacting 1D helical channel
in both the loop and Josephson junction geometries
within the Tomonaga-Luttinger liquid theory.
The expressions of the Hamiltonian and electron field are presented
in a bosonized form.
In Sect.~3, we calculate the local density of states
and discuss its characteristic behaviors.
The last section is devoted to the summary.
We set $\hbar = 1$ throughout this paper.

\section{Bosonized Model}

We focus on a weakly interacting 1D helical channel consisting of
right-going up-spin electrons and left-going down-spin electrons.
Before presenting its bosonized description,
let us briefly summarize the energy spectrum of a 1D helical channel
in the non-interacting limit.
Hereafter, $\sigma$ is used to denote the spin directions:
$\sigma = \uparrow$ for up spin and $\sigma = \downarrow$ for down spin,
and the sign function $s_{\sigma}$ is defined as
$s_{\sigma} = +$ ($-$) for $\sigma = \uparrow$ ($\downarrow$).

\begin{figure}[btp]
\begin{center}
\includegraphics[height=1.8cm]{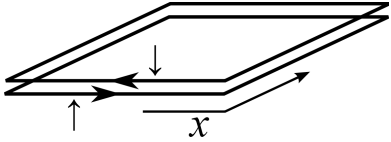}
\end{center}
\caption{Schematic of the 1D helical channel consisting of
right-going up-spin branch and left-going down-spin branch.
}
\end{figure}
Let us first consider the 1D helical channel of length $L$
in the loop geometry (see Fig.~1).
We assume that the loop encircles a magnetic flux $\phi$,
which induces the Aharonov-Bohm phase
\begin{align}
  \chi \equiv 2\pi\frac{\phi}{\phi_{0}} ,
\end{align}
where $\phi_{0}=2\pi/e$ is the flux quantum.
In the absence of electron-electron interactions,
the electrons in this system obey the following eigenvalue equation:
\begin{align}
   v_{\rm F}\bigl[ s_{\sigma}\left(-i\partial_{x}+eA_{x}\right)
                  -k_{\rm F} \bigr] u_{\sigma}(x)
   = \epsilon u_{\sigma}(x) ,
\end{align}
where $v_{\rm F}$ and $k_{\rm F}$ respectively are the Fermi velocity and
Fermi wave number,
and $A_{x}$ is the tangential component of the vector potential.
Note that the wave function satisfies the anti-periodic boundary condition
$u_{\sigma}(x)=-u_{\sigma}(x+L)$ in the helical channel
owing to the rotation of the spin-quantization axis.~\cite{takane-imura}
If $k_{\rm F}L=2\pi\times{\rm integer}$ is assumed for simplicity,
the energy eigenvalues for up-spin and down-spin electrons are given by
\begin{align}
      \label{eq:E_loop}
   \epsilon_{\sigma}
   = v_{\rm F}\left(q+\frac{\pi+s_{\sigma}\chi}{L}\right) ,
\end{align}
where $q = 2\pi n/L$ ($n = 0,\pm 1,\pm 2,\dots$).
The factor $\pi/L$ reflects the anti-periodic boundary condition.

We turn to the case of the Josephson junction,
in which the 1D helical channel is effectively confined
in the region of length $L$ by two superconductors (see Fig.~2),
where the right (left) superconductor occupies
the region of $x \ge L$ ($x \le 0$).
The electrons in this system
obey the following eigenvalue equation:~\cite{de-gennes}
\begin{align}
 & \left(\begin{array}{cc}
           v_{\rm F}\left(-i s_{\sigma}\partial_{x}-k_{\rm F}\right)
           & \Delta(x)e^{i\chi(x)} \\
           \Delta(x)e^{-i\chi(x)}
           & v_{\rm F}\left(i s_{\sigma}\partial_{x}+k_{\rm F}\right)
         \end{array}
   \right)
   \left(\begin{array}{c}
           u_{\sigma}(x) \\
           v_{\sigma}(x)
         \end{array}
   \right)
           \nonumber \\
 & \hspace{10mm}
 = \epsilon
   \left(\begin{array}{c}
           u_{\sigma}(x) \\
           v_{\sigma}(x)
         \end{array}
   \right) ,
\end{align}
where $\Delta(x)$ and $\chi(x)$, respectively, are the magnitude and phase
of the pair potential in the superconductors.
We assume that $\Delta(x)$ has a constant value, $\Delta_{0}$,
in the region occupied by the right or left superconductor and
vanishes in the region of $L > x > 0$,
and that $\chi(x)$ is equal to $\chi_{\rm R}$ ($\chi_{\rm L}$)
in the right (left) superconductor.
In the low-energy regime of $|\epsilon| < \Delta_{0}$,
the energy eigenvalues for up-spin and down-spin electrons are given by
\begin{align}
     \label{eq:E_JJ}
   \epsilon_{\sigma}
   = v_{\rm F}\left(q+\frac{\pi+s_{\sigma}\chi}{2L_{+}}\right) ,
\end{align}
where $q = \pi n/L_{+}$ ($n = 0,\pm 1,\pm 2,\dots$),
$L_{+} = L+\xi$ with $\xi \equiv v/\Delta_{0}$ being the coherence length,
and $\chi$ is the phase difference defined by
$\chi \equiv \chi_{\rm R} -\chi_{\rm L}$.
The effective length $L_{+}$ for electrons is slightly enlarged by
the phase shift induced in Andreev reflection processes.~\cite{takane1,takane2}
\begin{figure}[btp]
\begin{center}
\includegraphics[height=1.8cm]{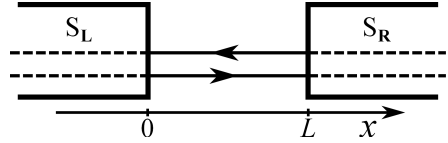}
\end{center}
\caption{Schematic of the 1D helical channel on which
two superconductors separated by $L$ is deposited.
}
\end{figure}

Now, we present a bosonized description of a weakly interacting 1D helical
channel in the two cases of the loop and Josephson junction geometries.
We describe the effect of electron-electron interactions
within the framework of the Tomonaga-Luttinger liquid.
Generally, the bosonized Hamiltonian $H$ is decomposed into
$H=H_{\rm Z}+H_{\rm NZ}$, where $H_{\rm Z}$ and $H_{\rm NZ}$
respectively describe the zero and non-zero modes.~\cite{haldane}
The electron field $\psi_{\sigma}(x)$ is expressed 
as~\cite{luther,suzumura1,loss}
\begin{align}
      \label{eq:psi-bosonized}
 \psi_{\sigma}(x)
 = \frac{1}{\sqrt{2\pi \alpha}}e^{is_\sigma k_{\rm F}x+i\theta_{\sigma}(x)} ,
\end{align}
where $\alpha$ is a short distance cutoff
and $\theta_{\sigma}(x)$ is the phase field.
With the explicit expression of $\theta_{\sigma}(x)$ given below,
we can show that the electron field satisfies the anti-commutation relation
\begin{align}
 \psi_{\sigma}(x)\psi_{\sigma'}^{\dagger}(x')
 + \psi_{\sigma'}^{\dagger}(x')\psi_{\sigma}(x)
 = \delta_{\sigma,\sigma'} \delta\left(x-x'\right)
\end{align}
for $L \ge x, x' \ge 0$.
The phase field is also decomposed into
the zero and non-zero mode components as follows:
\begin{align}
      \label{eq:theta-decomp}
   \theta_{\sigma}(x)
   = \theta_{\sigma}^{0}(x)
    + \frac{1}{2}\left(s_{\sigma}\theta_{+}(x)+\theta_{-}(x) \right) ,
\end{align}
where $\theta_{\sigma}^{0}(x)$ is the zero-mode component and
$\theta_{\pm}(x)$ describes the non-zero modes.
The explicit forms of the Hamiltonian and the phase field
necessarily reflect its geometry,
so we separately treat the two cases below.
However, the correlation exponent $K$ and the renormalized velocity $v$
are commonly used in both the cases,
where $K$ characterizes the strength of interactions and
$K<1$ in the ordinary case of repulsive interaction.
Note that $K = 1$ and $v = v_{\rm F}$ in the non-interacting limit.
For convenience, we here define $\gamma_{\pm}$ as
\begin{align}
  \gamma_{\pm} = \frac{1}{2}\left(\frac{1}{K} \pm K\right) .
\end{align}

\subsection{Loop geometry}

With the parameters $v$ and $K$ introduced above,
$H_{\rm Z}$ and $H_{\rm NZ}$ are respectively given by~\cite{haldane,loss}
\begin{align}
   H_{\rm Z}
 & = \frac{\pi v}{2L}
     \left(\frac{1}{K}M^2+K\left(J+\frac{\chi}{\pi}\right)^2 \right) ,
   \\
   H_{\rm NZ}
 & = \sum_{q>0}vq\left(\beta_{q}^{\dagger}\beta_{q}
                       + \beta_{-q}^{\dagger}\beta_{-q} \right) ,
\end{align}
where $J$ and $M$ are the winding numbers satisfying the constraint that
their sum must be an even integer,~\cite{loss} and $\beta_{q}$
($\beta_{q}^{\dagger}$) is the boson annihilation (creation) operator
with $q = 2\pi n/L$ ($n = 1,2,3,\dots$).
The zero-mode component of the phase field is given by
\begin{align}
  \theta_{\sigma}^{0}(x)
  = \vartheta + s_{\sigma}\varphi
    + \frac{\pi}{L}x\left(J+\frac{\chi}{\pi}\right)
    + s_{\sigma}\frac{\pi}{L}\left(x+\frac{L}{2}\right)M ,
\end{align}
with $[J,\varphi]=[M,\vartheta]=i$,
while the non-zero-mode component is given by
\begin{align}
  \theta_{+}(x)
  & = i\sqrt{K}\sum_{q>0}\sqrt{\frac{2\pi}{Lq}}e^{-\alpha q/2}
      \Bigl[ e^{-iqx}\left(\beta_{q}^{\dagger}+\beta_{-q}\right)
     \nonumber \\
  & \hspace{35mm}
             - e^{iqx}\left(\beta_{q}+\beta_{-q}^{\dagger}\right)
      \Bigr] ,
     \\
  \theta_{-}(x)
  & = \frac{i}{\sqrt{K}}\sum_{q>0}\sqrt{\frac{2\pi}{Lq}}e^{-\alpha q/2}
      \Bigl[ e^{-iqx}\left(\beta_{q}^{\dagger}-\beta_{-q}\right)
     \nonumber \\
  & \hspace{35mm}
             - e^{iqx}\left(\beta_{q}-\beta_{-q}^{\dagger}\right)
      \Big] .
\end{align}

\subsection{Josephson junction}

The bosonized description of the Josephson junction of
a spin-full Tomonaga-Luttinger liquid has been presented
by several authors.~\cite{maslov,takane1,affleck,takane2,takane3}
Applying the prescription to the helical case,
we find that~\cite{barbarino}
\begin{align}
   H_{\rm Z}
 & = \frac{\pi vK}{2L_{+}}
     \left(N+\frac{\chi}{2\pi}\right)^2 ,
   \\
   H_{\rm NZ}
 & = \sum_{q>0}vq \beta_{q}^{\dagger}\beta_{q} ,
\end{align}
where $q = \pi n/L_{+}$ ($n = 1,2,3,\cdots$).
The zero-mode component of the phase field is given by
\begin{align}
  \theta_{\sigma}^{0}(x)
  = s_{\sigma}\varphi
    + \frac{\pi}{L_{+}}\left(x + \frac{\xi}{2}\right)
      \left(N+\frac{\chi}{2\pi}\right) ,
\end{align}
with $[N,\varphi]=i$, while the non-zero-mode component is given by
\begin{align}
  \theta_{+}(x)
  & = i\sqrt{K}\sum_{q>0}\sqrt{\frac{4\pi}{L_{+}q}}e^{-\alpha q/2}
      \cos q\left(x+\frac{\xi}{2}\right)
      \left(\beta_{q}^{\dagger}-\beta_{q}\right) ,
       \\
  \theta_{-}(x)
  & = \frac{1}{\sqrt{K}}\sum_{q>0}\sqrt{\frac{4\pi}{L_{+}q}}e^{-\alpha q/2}
      \sin q\left(x+\frac{\xi}{2}\right)
      \left(\beta_{q}^{\dagger}+\beta_{q}\right) .
\end{align}
In contrast to the loop geometry case, $\theta_{\pm}(x)$ has
a nontrivial spatial dependence as a consequence of the Andreev reflection.
For example, the fluctuation of $\theta_{-}$ is strongly suppressed near
both ends of the system (i.e., $x = 0$ and $L$).~\cite{takane2,takane3}
This results in anomalous $x$- and $\omega$-dependences
of the local density of states.

\section{Local Density of States}

To obtain an analytical expression of the local density of states
at zero temperature,
it is convenient to introduce the retarded Green's function defined by
\begin{align}
  G_{\sigma}^{\rm R}(x,x';t)
  & = -i\Theta(t)
       \Bigl[ \langle \psi_{\sigma}(xt)\psi_{\sigma}^{\dagger}(x'0) \rangle
          \nonumber \\
  & \hspace{20mm}
           + \langle \psi_{\sigma}^{\dagger}(x'0)\psi_{\sigma}(xt) \rangle
      \Bigr] ,
\end{align}
where $\Theta(t)$ is the Heaviside step function, $\langle \cdots \rangle$
represents the average in the ground state, and
\begin{align}
  \psi_{\sigma}(xt)
  = e^{iHt}\psi_{\sigma}(x)e^{-iHt} .
\end{align}
Then, the local density of states at $x$ is expressed as
\begin{align}
     \label{eq:dos-def}
  D_{\sigma}(x,\omega)
   = -\frac{1}{\pi}{\rm Im}
      \left\{\int_{-\infty}^{\infty}\!dt \,
             e^{i\omega t}G_{\sigma}^{\rm R}(x,x;t)
      \right\} .
\end{align}
Below, we separately treat the cases of
loop and Josephson junction geometries.

\subsection{Loop geometry}

Using the bosonized expressions for the Hamiltonian and electron field,
we obtain the retarded Green's function as
\begin{align}
  G_{\sigma}^{\rm R}(x,x;t)
  & = -i\Theta(t)\frac{1}{2\pi \alpha}
      \left(\frac{2\pi \alpha}{L}\right)^{\gamma_{+}}
      e^{-is_{\sigma}\frac{vK\chi}{L}t}
        \nonumber \\
  & \hspace{-10mm}
      \times
      \left( \frac{e^{-i\frac{\pi v\gamma_{+}}{L}t}}
                  {\left(1-e^{-\frac{2\pi}{L}(\alpha + ivt)}
                   \right)^{\gamma_{+}}}
             + {\rm c.c.}
      \right) .
\end{align}
In deriving this, it is implicitly assumed
that $J = M = 0$ in the ground state.
This is justified when $|\chi|<\pi/2$.
It is clear that $G_{\sigma}^{\rm R}(x,x;t)$ has no spatial dependence
reflecting the translational invariance of the system.
The substitution of this into Eq.~(\ref{eq:dos-def}) straightforwardly yields
\begin{align}
      \label{eq:dos-loop_1}
  D_{\sigma}(x, \omega)
  & = \frac{1}{4\pi^2 \alpha}
      \left(\frac{2\pi \alpha}{L}\right)^{\gamma_{+}}
      \int_{-\infty}^{\infty}\! dt \, 
      e^{i(\omega-s_{\sigma}\frac{vK\chi}{L})t}
        \nonumber \\
  & \hspace{-0mm}
      \times
      \left( \frac{e^{-i\frac{\pi v\gamma_{+}}{L}t}}
                  {\left(1-e^{-\frac{2\pi}{L}(\alpha + ivt)}
                   \right)^{\gamma_{+}}}
             + {\rm c.c.}
      \right) .
\end{align}

To simplify the expression of $D_{\sigma}(x, \omega)$,
we employ the following binomial expansion:
\begin{align}
      \label{eq:def-bi_expan}
  \frac{1}{\left(1-e^{-\frac{2\pi}{L}(\alpha \pm ivt)}\right)^{\gamma_{+}}}
  = \sum_{n=0}^{\infty}
    a_{n}(\gamma_{+})e^{-\frac{2\pi}{L}n(\alpha \pm ivt)}
\end{align}
with $a_{0}(\gamma_{+}) = 1$ and
\begin{align}
      \label{eq:def-bi_c}
  a_{n}(\gamma_{+})
  = \frac{\gamma_{+}(\gamma_{+}+1)\cdots(\gamma_{+}+n-1)}{n!}
\end{align}
for $n \ge 1$.
Substituting this into Eq.~(\ref{eq:dos-loop_1}) and carrying out
the integration over $t$, we obtain
\begin{align}
      \label{eq:dos-loop_2}
  D_{\sigma}(x, \omega)
  & = \frac{1}{L}\left(\frac{2\pi \alpha}{L}\right)^{\gamma_{+}-1}
      \sum_{n=0}^{\infty}
      a_{n}(\gamma_{+})
        \nonumber \\
  & \hspace{-10mm}
      \times
      \biggl[ \delta\left( \omega-\frac{\pi v}{L}
                           \left( 2n+\gamma_{+}
                                  +s_{\sigma}\frac{K\chi}{\pi} \right)
                   \right)
        \nonumber \\
  & \hspace{5mm}
           + \delta\left( \omega+\frac{\pi v}{L}
                          \left( 2n+\gamma_{+}
                                 -s_{\sigma}\frac{K\chi}{\pi} \right)
                   \right)
      \biggr] .
\end{align}

Equation~(\ref{eq:dos-loop_2}) indicates that the local density of states
consists of equally spaced spikes of the $\delta$-function type.
We easily see that, in the non-interacting limit of $K=1$ and $v=v_{\rm F}$,
the location of each spike exactly corresponds to Eq.~(\ref{eq:E_loop}).
It also indicates that the height of the $n$th spike is characterized
by the corresponding binomial coefficient $a_{n}(\gamma_{+})$.
That is, the effect of interactions appears in the relative height
of succeeding spikes.
Figure~3 schematically shows the local density of states with $K = 0.7$,
where the spikes up to $n = 20$ on the side of $\omega > 0$ are shown.
The $n$th bar represents the relative height of the $n$th spike
normalized by the $0$th spike.
\begin{figure}[btp]
\begin{center}
\includegraphics[height=4.0cm]{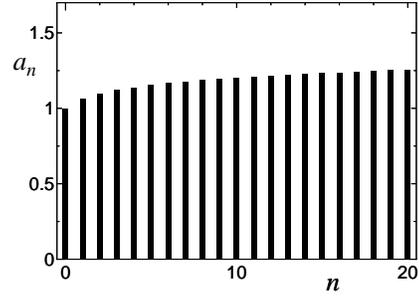}
\end{center}
\caption{Schematic of the local density of states in the loop geometry
on the side of $\omega > 0$ with $K = 0.7$.
Each bar represents the relative height of the corresponding spike
normalized by that of the $0$th spike.
}
\end{figure}
As seen in Fig.~3, the height of spikes decreases with decreasing $n$
according to Eq.~(\ref{eq:def-bi_c})
with the fact that $\gamma_{+} > 1$ regardless of $K$.
The behavior similar to this can also be observed
in the limit of $L \to \infty$.
Indeed, we can show that in the large-$L$ limit,
Eq.~(\ref{eq:dos-loop_2}) is reduced to
\begin{align}
     \label{eq:dos-L-to-inf}
  D_{\sigma}(x, \omega)
  \to \frac{1}{2\pi v\Gamma(\gamma_{+})}
      \left(\frac{\alpha|\omega|}{v}\right)^{\gamma_{+}-1} .
\end{align}
This indicates that the density of states is suppressed
in the low-energy limit of $\omega \to 0$.~\cite{luther,suzumura2}

\subsection{Josephson junction}

Repeating the procedure carried out in the loop geometry case,
we obtain the retarded Green's function for $|\chi|<\pi$ as
\begin{align}
  G_{\sigma}^{\rm R}(x,x;t)
  & = -i\Theta(t)\frac{1}{2\pi \alpha}
      \left(\frac{\pi \alpha}{L_{+}}\right)^{\gamma_{+}}
      e^{-is_{\sigma}\frac{vK\chi}{2L_{+}}t}
        \nonumber \\
  & \hspace{-15mm}
      \times
      \left( \frac{e^{-i\frac{\pi vK}{2L_{+}}t}}
                  {\left(1-e^{-\frac{\pi}{L_{+}}(\alpha + ivt)}
                   \right)^{\gamma_{+}}}\Omega(x,t)
             + {\rm c.c.}
      \right)
\end{align}
with
\begin{align}
  \Omega(x,t)
  & = \left( \frac{1-e^{-\frac{\pi}{L_{+}}(\alpha + i(vt-2x-\xi))}}
                  {1-e^{-\frac{\pi}{L_{+}}(\alpha - i(2x+\xi))}}
      \right)^{\frac{1}{2}\gamma_{-}}
        \nonumber \\
  & \hspace{5mm}
      \times
      \left( \frac{1-e^{-\frac{\pi}{L_{+}}(\alpha + i(vt+2x+\xi))}}
                  {1-e^{-\frac{\pi}{L_{+}}(\alpha + i(2x+\xi))}}
      \right)^{\frac{1}{2}\gamma_{-}} .
\end{align}
In this case, $G_{\sigma}^{\rm R}(x,x;t)$ has a spatial dependence
reflecting the inhomogeneity of the non-zero mode component of the phase field.
Substituting this into Eq.~(\ref{eq:dos-def}), we straightforwardly obtain
\begin{align}
      \label{eq:dos-Josephson_1}
  D_{\sigma}(x, \omega)
  & = \frac{1}{4\pi^2 \alpha}
      \left(\frac{\pi \alpha}{L_{+}}\right)^{\gamma_{+}}
      \int_{-\infty}^{\infty}\! dt \,
      e^{i(\omega-s_{\sigma}\frac{vK\chi}{2L_{+}})t}
        \nonumber \\
  & \hspace{-0mm}
      \times
      \left( \frac{e^{-i\frac{\pi vK}{2L_{+}}t}}
                  {\left(1-e^{-\frac{\pi}{L_{+}}(\alpha + ivt)}
                   \right)^{\gamma_{+}}}\Omega(x,t)
             + {\rm c.c.}
      \right) .
\end{align}

Even in this case, we can derive a general expression for
$D_{\sigma}(x, \omega)$ using binomial expansions.
However, as the resulting expression is slightly lengthy and complicated,
we restrict our attention to the behavior of $D_{\sigma}(x, \omega)$
at the end point of $x=0$ (or equivalently $x = L$),
at the midpoint of $x = L/2$, and at quarter point of $x = L/4$.

\subsubsection{Local density of states at $x=0$}

At $x=0$, Eq.~(\ref{eq:dos-Josephson_1}) is simplified to
\begin{align}
      \label{eq:dos-Josephson_2}
  D_{\sigma}(0, \omega)
  & = \frac{1}{4\pi^2 \alpha}
      \left(\frac{\pi \alpha}{L_{+}}\right)^K
      \int_{-\infty}^{\infty}\! dt \,
      e^{i(\omega-s_{\sigma}\frac{vK\chi}{2L_{+}})t}
        \nonumber \\
  & \hspace{-0mm}
      \times
      \left( \frac{e^{-i\frac{\pi vK}{2L_{+}}t}}
                  {\left(1-e^{-\frac{\pi}{L_{+}}(\alpha + ivt)}
                   \right)^K}
             + {\rm c.c.}
      \right) .
\end{align}
Applying the procedure used to derive Eq.~(\ref{eq:dos-loop_2}),
we find that
\begin{align}
      \label{eq:dos-Josephson_3}
  D_{\sigma}(0, \omega)
  & = \frac{1}{2L_{+}}\left(\frac{\pi\alpha}{L_{+}}\right)^{K-1}
      \sum_{n=0}^{\infty} a_{n}(K)
        \nonumber \\
  & \hspace{-10mm}
      \times
      \biggl[ \delta\left(\omega-\frac{\pi v}{L_{+}}
                          \left(n+\frac{K}{2}
                          +s_{\sigma}\frac{K\chi}{2\pi}\right)
                    \right)
        \nonumber \\
  & \hspace{5mm}
           + \delta\left(\omega+\frac{\pi v}{L_{+}}
                         \left(n+\frac{K}{2}
                         -s_{\sigma}\frac{K\chi}{2\pi} \right)
                   \right)
      \biggr] ,
\end{align}
where the binomial coefficient is defined in Eq.~(\ref{eq:def-bi_c}).

Similar to the loop case,
the local density of states consists of equally spaced spikes
and, in the non-interacting limit, the location of each spike
exactly corresponds to Eq.~(\ref{eq:E_JJ}).
Again, the height of spikes is modified by electron-electron interactions.
However, in this case, it is determined by $K$ instead of $\gamma_{+}$.
This should be attributed to the fact that the fluctuation of $\theta_{-}(x)$
is significantly suppressed at the ends of the system.
Figure~4 schematically shows the local density of states
with $K = 0.7$, where spikes up to $n = 20$ are shown.
This clearly indicates that the height of spikes increases with decreasing $n$
in contrast to the loop geometry case, and that
its variation is much more pronounced than that in the loop geometry case.
\begin{figure}[btp]
\begin{center}
\includegraphics[height=4.0cm]{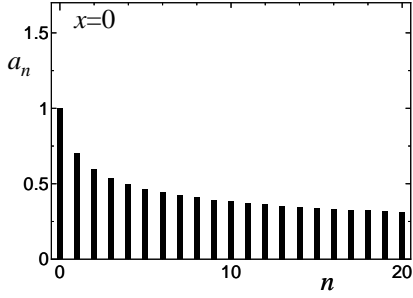}
\end{center}
\caption{Schematic of the local density of states at the end point
on the side of $\omega > 0$ with $K = 0.7$.
Each bar represents the relative height of the corresponding spike
normalized by that of the $0$th spike.
}
\end{figure}
This behavior manifests itself as long as $K < 1$,
which is ensured in the ordinary case of repulsive interaction.
The behavior similar to this can also be observed
in the limit of $L \to \infty$.
Indeed, we can show that in the large-$L$ limit,
Eq.~(\ref{eq:dos-Josephson_3}) is reduced to
\begin{align}
  D_{\sigma}(0, \omega)
  \to \frac{1}{2\pi v\Gamma(K)}\left(\frac{\alpha|\omega|}{v}\right)^{K-1} .
\end{align}
This indicates that the density of states is enhanced
in the low-energy limit of $\omega \to 0$.~\cite{winkelholz,tilahun}

\subsubsection{Local density of states at $x=L/2$ and $L/4$}

At $x = L/2$, Eq.~(\ref{eq:dos-Josephson_1}) is simplified to
\begin{align}
      \label{eq:dos-Josephson_4}
  D_{\sigma}(L/2, \omega)
  & = \frac{1}{4\pi^2 \alpha}
      \left(\frac{1}{2}\right)^{\gamma_{-}}
      \left(\frac{\pi \alpha}{L_{+}}\right)^{\gamma_{+}}
        \nonumber \\
  & \hspace{-15mm}
      \times
      \int_{-\infty}^{\infty}\! dt \,
      e^{i(\omega-s_{\sigma}\frac{vK\chi}{2L_{+}})t}
        \nonumber \\
  & \hspace{-15mm}
      \times
      \left( \frac{e^{-i\frac{\pi vK}{2L_{+}}t}
                   \left(1+e^{-\frac{\pi}{L_{+}}(\alpha + ivt)}
                   \right)^{\gamma_{-}}}
                  {\left(1-e^{-\frac{\pi}{L_{+}}(\alpha + ivt)}
                   \right)^{\gamma_{+}}}
             + {\rm c.c.}
      \right) .
\end{align}
The factor in the bracket can be expanded in a double series as
\begin{align}
      \label{eq:doubl-series}
  & \frac{\left(1+e^{-\frac{\pi}{L_{+}}(\alpha \pm ivt)}
               \right)^{\gamma_{-}}}
         {\left(1-e^{-\frac{\pi}{L_{+}}(\alpha \pm ivt)}
               \right)^{\gamma_{+}}}
           \nonumber \\
  & \hspace{-5mm}
  = \sum_{l=0}^{\infty}\sum_{m=0}^{\infty}
    a_{l}(\gamma_{+})b_{m}(\gamma_{-})
    e^{-\frac{\pi}{L_{+}}(l+m)(\alpha \pm ivt)} ,
\end{align}
where $b_{0}(\gamma_{-})=1$ and
\begin{align}
      \label{eq:def-bi_d}
  b_{m}(\gamma_{-})
  = \frac{\gamma_{-}(\gamma_{-}-1)\cdots(\gamma_{-}-m+1)}{m!}
\end{align}
for $m \ge 1$.
Substituting Eq.~(\ref{eq:doubl-series}) into Eq.~(\ref{eq:dos-Josephson_4})
and then carrying out the integration, we find that
\begin{align}
      \label{eq:dos-Josephson_5}
  D_{\sigma}(L/2, \omega)
  & = \frac{1}{2L_{+}}
      \left(\frac{1}{2}\right)^{\gamma_{-}}
      \left(\frac{\pi\alpha}{L_{+}}\right)^{\gamma_{+}-1}
      \sum_{n=0}^{\infty}c_{n}
        \nonumber \\
  & \hspace{-15mm}
      \times
      \biggl[ \delta\left(\omega-\frac{\pi v}{L_{+}}
                          \left(n+\frac{K}{2}
                          +s_{\sigma}\frac{K\chi}{2\pi}\right)
                    \right)
        \nonumber \\
  & \hspace{-5mm}
           + \delta\left(\omega+\frac{\pi v}{L_{+}}
                         \left(n+\frac{K}{2}
                         -s_{\sigma}\frac{K\chi}{2\pi}\right)
                   \right)
      \biggr] ,
\end{align}
where $c_{n}$, which determines the height of the $n$th spike, is given by
\begin{align}
  c_{n} = \sum_{m=0}^{n} a_{n-m}(\gamma_{+})b_{m}(\gamma_{-}) .
\end{align}
The coefficients up to $n = 4$ are given as follows:
\begin{align}
  c_{0}
  & = 1 ,
     \\
  c_{1}
  & = \frac{1}{K} ,
     \\
  c_{2}
  & = \frac{1}{2}\left(\frac{1}{K^2}+K\right) ,
     \\
  c_{3}
  & = \frac{1}{6}\left(\frac{1}{K^2}+\frac{2}{K}+3\right) ,
     \\
  c_{4}
  & = \frac{1}{24}
      \left(\frac{1}{K^4}+\frac{8}{K^2}+\frac{6}{K}+6K+3K^2\right) .
\end{align}

The derivation of the local density of states at $x = L/4$ is similar to
that at $x = L/2$, so we present only the final result:
\begin{align}
      \label{eq:dos-Josephson_6}
  D_{\sigma}(L/4, \omega)
  & = \frac{1}{2L_{+}}
      \left(\frac{1}{2}\right)^{\frac{\gamma_{-}}{2}}
      \left(\frac{\pi\alpha}{L_{+}}\right)^{\gamma_{+}-1}
      \sum_{n=0}^{\infty}d_{n}
        \nonumber \\
  & \hspace{-15mm}
      \times
      \biggl[ \delta\left(\omega-\frac{\pi v}{L_{+}}
                          \left(n+\frac{K}{2}
                          +s_{\sigma}\frac{K\chi}{2\pi}\right)
                    \right)
        \nonumber \\
  & \hspace{-5mm}
           + \delta\left(\omega+\frac{\pi v}{L_{+}}
                         \left(n+\frac{K}{2}
                         -s_{\sigma}\frac{K\chi}{2\pi}\right)
                   \right)
      \biggr] ,
\end{align}
where $d_{n}$, which determines the height of the $n$th spike, is given by
\begin{align}
  d_{n} = \sum_{m=0}^{\left[n/2\right]}
          a_{n-2m}(\gamma_{+})b_{m}(\gamma_{-}/2) .
\end{align}
The coefficients up to $n = 4$ are given as follows:
\begin{align}
  d_{0}
  & = 1 ,
     \\
  d_{1}
  & = \frac{1}{2}\left(\frac{1}{K}+K\right) ,
     \\
  d_{2}
  & = \frac{1}{8}\left(\frac{1}{K^2}+\frac{4}{K}+2+K^2\right) ,
     \\
  d_{3}
  & = \frac{1}{48}\left(\frac{1}{K^3}+\frac{12}{K^2}+\frac{11}{K}
                        +12+11K+K^3\right) ,
     \\
  d_{4}
  & = \frac{1}{384}
      \biggl(\frac{1}{K^4}+\frac{24}{K^3}+\frac{84}{K^2}+\frac{48}{K}+70
          \nonumber \\
  & \hspace{25mm}
            +120K+36K^2+K^4\biggr) .
\end{align}

Figures~5(a) and 5(b) respectively show the local density of states
at $x = L/2$ and $L/4$ with $K = 0.7$,
where spikes up to $n = 20$ are shown.
Note that the $n$-dependence of $c_{n}$ and $d_{n}$ is not
monotonic in contrast to the case at the end points.
This should be attributed to inhomogeneous spatial fluctuations
of the non-zero mode component $\theta_{\pm}(x)$ of the phase field.
We expect that such a non-monotonic $n$-dependence
is observed except in the vicinity of the end points.
\begin{figure}[btp]
\begin{center}
\includegraphics[height=4.0cm]{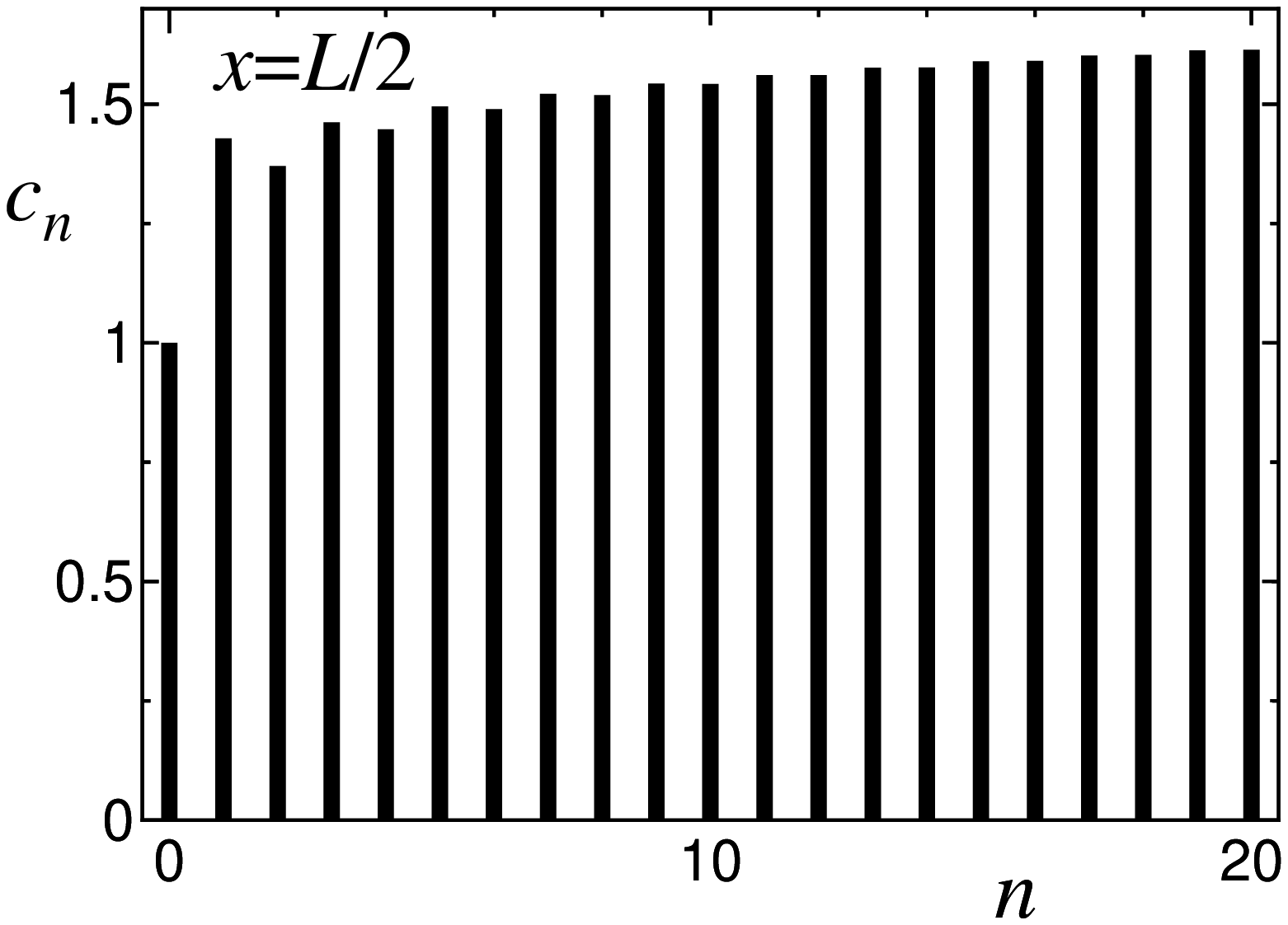}
\includegraphics[height=4.0cm]{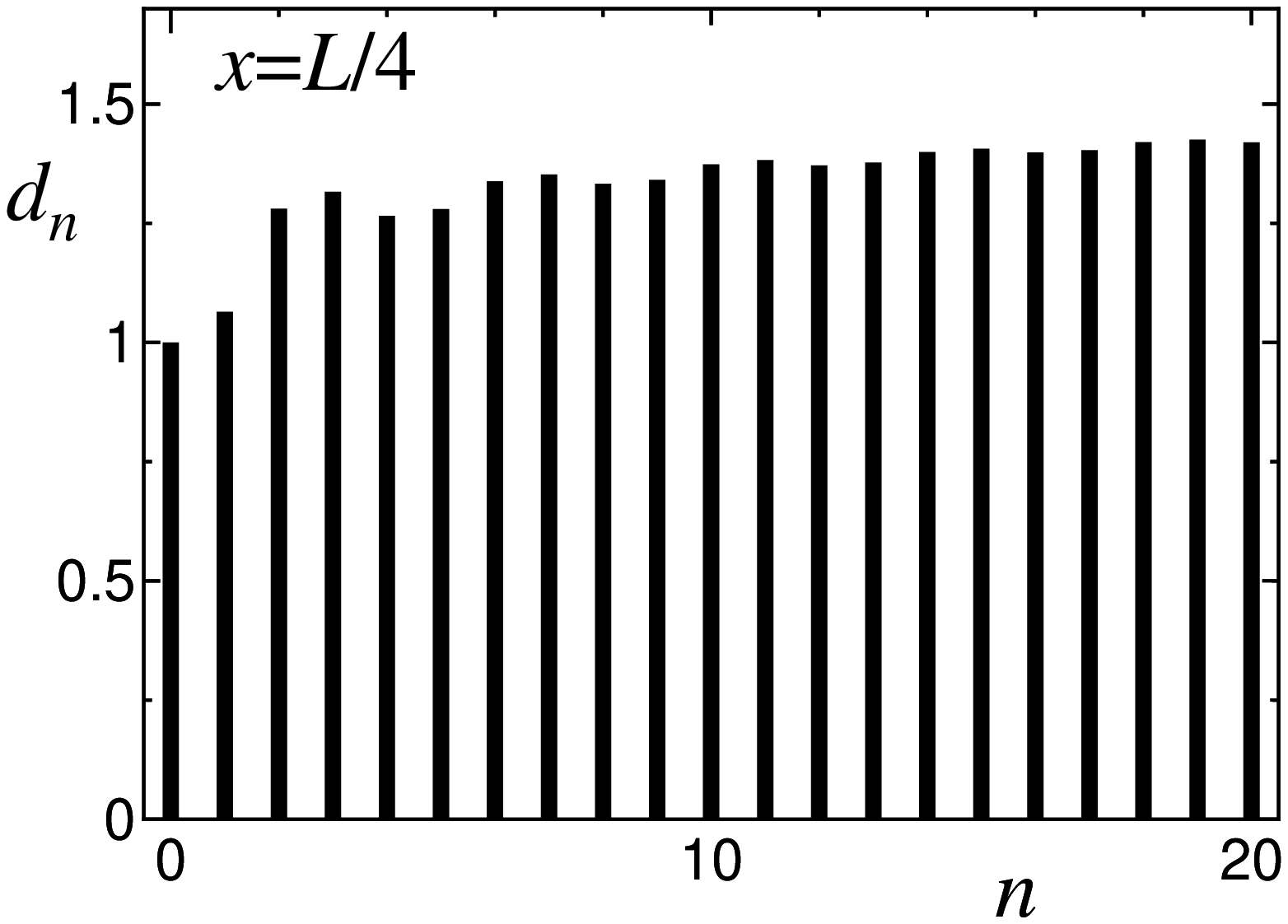}
\end{center}
\caption{Schematic of the local density of states at (a) $x = L/2$
and (b) $x = L/4$ on the side of $\omega > 0$ with $K = 0.7$.
Each bar represents the relative height of the corresponding spike
normalized by that of the $0$th spike.
}
\end{figure}

In the limit of $L \to \infty$, we expect that, since the effect of
superconductors plays no role except in the vicinity of the end points,
the local density of states becomes equivalent to the corresponding result,
Eq.~(\ref{eq:dos-L-to-inf}), in the loop geometry case,
and hence the non-monotonic $n$-dependence of peak heights disappears.
Indeed, we can show that Eqs.~(\ref{eq:dos-Josephson_5})
and (\ref{eq:dos-Josephson_6}) become equivalent to
Eq.~(\ref{eq:dos-L-to-inf}) in the large-$L$ limit.

\section{Summary}

The local density of states in a one-dimensional helical edge channel
is studied within the framework of a Tomonaga-Luttinger liquid
at zero temperature.
To observe the finite-size effect combined with
the effect of electron-electron interactions,
the two particular cases
of loop and Josephson junction geometries are considered.
The local density of states, as a function of the energy $\omega$, consists of
equally spaced spikes of the $\delta$-function type,
and electron-electron interactions modify their relative height.
It is shown that the height of spikes decreases as $\omega \to 0$
in the loop geometry case.
It is also shown that, in the Josephson junction case,
the behavior of the local density of states significantly depends
on the spatial position.
At the end points of the junction, the height increases as $\omega \to 0$
and its variation is more pronounced than that in the loop case,
while a non-monotonic $\omega$-dependence of spikes is found
away from the end points.

The experimental detection of these behaviors
using scanning tunneling microscopy is an interesting future challenge.
In principle, phenomena similar to these behaviors also appear
in a spin-full Tomonaga-Luttinger liquid.~\cite{winkelholz}
However, their detection is much more difficult in the spin-full case
than in the helical case because the spin-full Tomonaga-Luttinger liquid
is significantly affected by single-particle backward scattering
induced by disorder.

\section*{Acknowledgment}

This work is partially supported by a Grant-in-Aid for Scientific Research
(C) (No. 24540375).

\appendix


\begin{thebibliography}{99}
\bibitem{comment} Each spike at zero temperature is not smeared
even in the presence of interactions reflecting
the integrability of the Tomonaga-Luttinger model.
\end{thebibliography}
\end{document}